\documentclass{article}

\usepackage{arxiv}

\usepackage[utf8]{inputenc} 
\usepackage[T1]{fontenc}    
\usepackage[colorlinks, allcolors=blue]{hyperref}       
\usepackage{url}            
\usepackage{booktabs}       
\usepackage{amsfonts}       
\usepackage{nicefrac}       
\usepackage{microtype}      
\usepackage{lipsum}	    	
\usepackage{graphicx}
\usepackage{doi}
\usepackage{bm}
\usepackage{mathtools}
\usepackage[usenames,dvipsnames,table,modulo]{xcolor}
\usepackage[nolist]{acronym}
\usepackage{xspace}
\usepackage{lineno}
\usepackage{authblk}
\usepackage{multirow}
\usepackage[capitalise]{cleveref}
\usepackage{setspace} 
\newcommand{\hitnet}{HitNet\xspace}
\newcommand{\chargenet}{ChargeNet\xspace}
\newcommand{\perdom}{per-sensor\xspace}
\newcommand{\alldom}{all-sensor\xspace}
\newcommand{\ie}{\emph{i.e.}\xspace}
\newcommand{\eg}{\emph{e.g.}\xspace}
\newcommand{\params}{\ensuremath{\bm{\theta}}\xspace}
\newcommand{\obs}{\ensuremath{\bm{x}}\xspace}

\title{A flexible event reconstruction based on machine learning and likelihood principles}

\author[1]{Philipp Eller \thanks{\href{mailto:philipp.eller@tum.de}{philipp.eller@tum.de}}\ \ }
\author[3]{Aaron T. Fienberg}
\author[2]{Jan Weldert \thanks{\href{mailto:weldert@uni-mainz.de}{weldert@uni-mainz.de}}\ \ }
\author[3]{Garrett Wendel \thanks{\href{mailto:gmw5164@psu.edu}{gmw5164@psu.edu}}\ \ }
\author[2]{Sebastian B\"oser}
\author[3,4]{D. F. Cowen}
\affil[1]{Technical University Munich, Garching, Germany}
\affil[2]{Johannes Gutenberg University, Mainz, Germany}
\affil[3]{Department of Physics, The Pennsylvania State University, University Park, PA, 16802 USA}
\affil[4]{Department of Astronomy and Astrophysics, The Pennsylvania State University, University Park, PA, 16802 USA}

\renewcommand{\shorttitle}{}

\begin{document}
\maketitle

\begin{acronym}[TDMA]
\acro{BDT}{Boosted Decision Tree}
\acro{SK}{Super-Kamiokande}
\acro{HK}{Hyper-Kamiokande}
\acro{SM}{Standard Model}
\acro{BSM}{Beyond the Standard Model}
\acro{MC}{Monte Carlo}
\acro{ML}{machine learning}
\acro{MLP}{Multilayer Perceptron}
\acro{NMO}{Neutrino Mass Ordering}
\acro{CNN}{Convolutional Neural Network}
\acro{GNN}{Graph Neural Network}
\acro{GAN}{Generative Adversarial Network}
\acro{GPU}{Graphics Processing Unit}
\acro{CPU}{Central Processing Unit}
\acro{CC}{charged current}
\acro{NC}{neutral current}
\acro{DOM}{Digital Optical Module}
\acro{MTOM}{Muon Tomography Optical Module}
\acro{SiPM}{Silicon Photomultiplier}
\acro{PMT}{Photomultiplier Tube}
\acro{MCP}{Microchannel Plate}
\acro{DAQ}{Data Acquisition}
\acro{CR}{Cosmic Ray}
\acro{PMNS}{Pontecorvo-Maki-Nakagawa-Sakata}
\acro{MSW}{Mikheyev-Smirnov-Wolfenstein}
\acro{DUNE}{Deep Underground Neutrino Experiment}
\acro{JUNO}{Jiangmen Underground Neutrino Observatory}
\acro{FPGA}{Field-Programmable Gate Array}
\acro{IC}{IceCube}
\acro{DC}{DeepCore}
\acro{NN}{neural network}
\acro{ROC}{Receiver Operating Characteristics}
\acro{QE}{quantum efficiency}
\acro{BCE}{Binary Cross Entropy}
\acro{AUC}{Area Under the Curve}
\acro{MLE}{maximum likelihood estimator}
\acro{EML}{extended maximum likelihood}
\acro{PDF}{probability density function}
\acro{LLH}{log-likelihood}
\end{acronym}

\begin{abstract}

Event reconstruction is a central step in many particle physics experiments, turning detector observables into parameter estimates; for example estimating the energy of an interaction given the sensor readout of a detector.
A corresponding likelihood function is often intractable, and approximations need to be constructed.
In our work, we first show how the full likelihood for a many-sensor detector can be broken apart into smaller terms, and secondly how we can train neural networks to approximate all terms solely based on forward simulation.
Our technique results in a fast, flexible, and close-to-optimal surrogate model proportional to the likelihood and can be used in conjunction with standard inference techniques allowing for a consistent treatment of uncertainties.
We illustrate our technique for parameter inference in neutrino telescopes based on maximum likelihood and Bayesian posterior sampling. Given its great flexibility, we also showcase our method for geometry optimization enabling to learn optimal detector designs. Lastly, we apply our method to realistic simulation of a ton-scale water-based liquid scintillator detector.

\end{abstract}


\section{Introduction}

A crucial step in the processing chain of many particle physics experiments is the so-called ``reconstruction'' step, where the parameters of interest in individual interactions \params (\eg, energy, momentum, direction, vertex position) are estimated based on observed measurements \obs.
A popular statistical technique that results in desirable estimators is the maximum likelihood technique. 
The likelihood function defines the probabilistic distribution of observables \obs for a given \params, and for a fixed \obs the value $\bm{\hat{\theta}}$ that maximizes the likelihood is denoted the \ac{MLE}. 
The likelihood function can also be used to calculate uncertainties on \acp{MLE} (confidence level regions), and likelihood ratio tests can be used to compare hypotheses. In a Bayesian context, the likelihood can be used to evaluate posterior probabilities over desired parameters.

However, creating the correct likelihood function can be a challenge or may even be intractable, as is often the case for many modern physics experiments (e.g. the IceCube Neutrino Observatory~\cite{IceCube:2016zyt}). In these instances, one needs to resort to approximations and simplifications to compute a likelihood.
The \ac{ML} technique of \acp{NN} offers a framework for universal function approximation, making it a suitable candidate for surrogate models. It is in principle possible to train \acp{NN} to directly approximate the simulation chain of an experiment, and use them as the expectation values to formulate a likelihood.
In this paper, however, we follow a likelihood-free technique in that we only require simulated data as input without an explicit assumption of the likelihood function.
We illustrate our technique, showing how the problem can be split up into smaller problems under the assumption of independence. 
We apply our method for event reconstruction to water-based Cherenkov~\cite{Cherenkov:1934ilx} neutrino detectors (see, \eg,~\cite{SNO:1999crp, Super-Kamiokande:2002weg, IceCube:2006tjp, IceCube:2016zyt, JUNO:2022hxd}), comparing its performance to truth level information for a realistic but very simplified ``toy'' experiment. Finally, we apply our method for the optimization of the design of an actual neutrino detector using a realistic simulation.

\section{Method}

Particle physics detectors typically comprise a large number of discrete sensors or channels. For example, a neutrino detector can feature thousands of photosensors such as photomultiplier tubes (PMTs), or a tracking detector can feature thousands of readout channels. Henceforth, we will simply use the term sensors for simplicity.
A detected particle interaction will generally lead to observed activity in only a fraction of these sensors. Additionally, a single sensor may yield multiple observations, here denoted as ``hits.'' For a single detected particle interaction---an ``event''---the complete observation $\obs$ takes the form of a variable-length collection of hits to which each sensor contributes $\ge\! 0$ times. (The specific content of a hit will vary according to the particle detector in question, but a simple example is the case where each hit contains a sensor label and a pulse time.) The size of the hit collection, $N$, or total number of hits, is an important component of the observation that carries information about the parameters of the particle interaction, \params. Upon repeated observations of a given \params (\eg, by simulating the same event multiple times), \obs will vary according to statistical fluctuations. The task of reconstruction is to infer \params given \obs. 

We represent an observation \obs by a set of hits, each of which has a sensor label and measurements of arrival time, and brightness or ``light level.'' The sensor label provides the sensor's position relative to other sensors.  The time represents the photon(s) arrival time and the light level the number of photons in the hit that arrive within the time resolution of the sensor. As sensors typically convert the light level into an electrical pulse with proportionate integrated charge, we will refer to the ``charge'' of a hit.

\begin{figure}[h]
    \centering
    \includegraphics[width=0.5\textwidth]{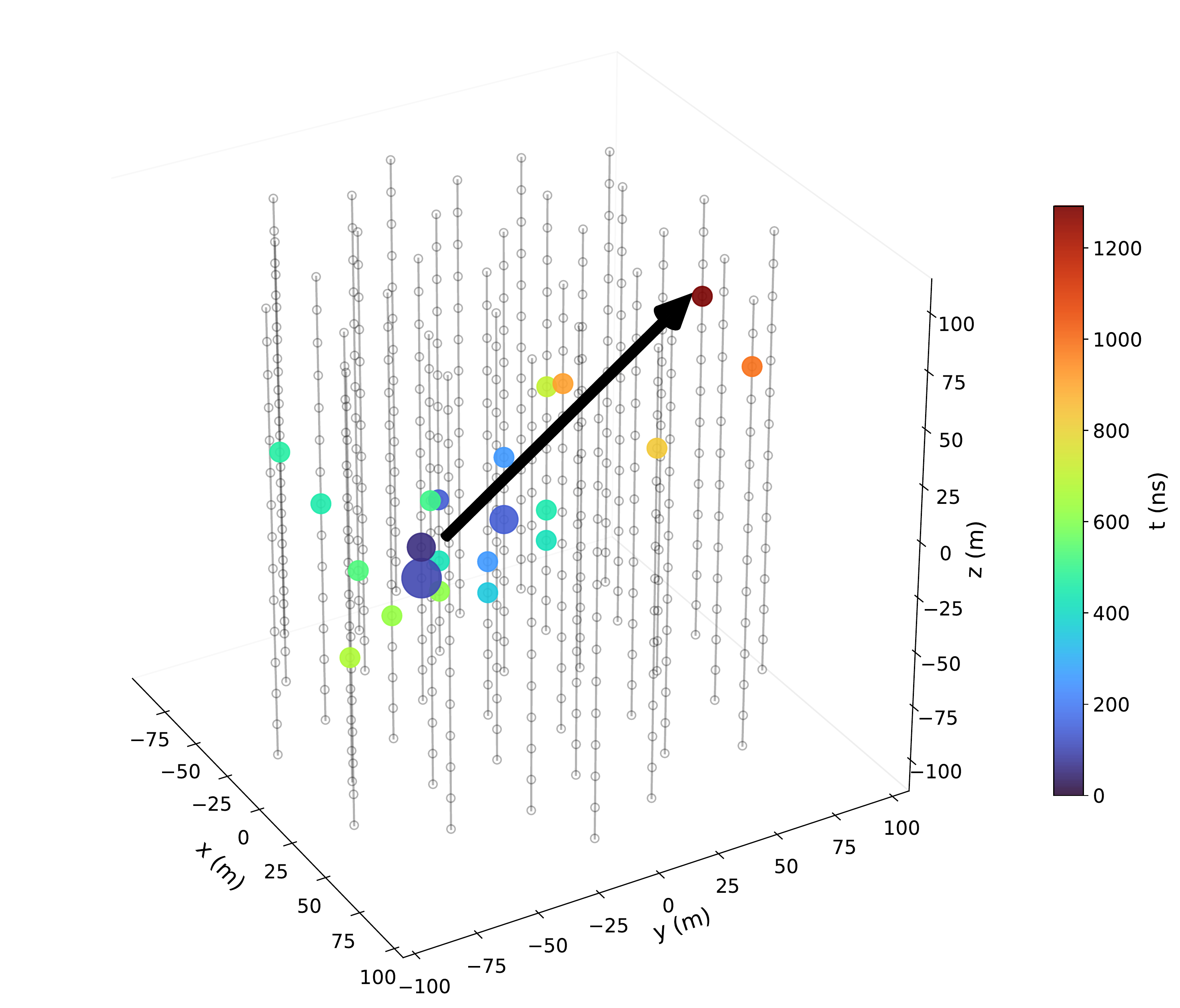}
    \caption{Display of an example track-like event in a toy neutrino telescope experiment. The gray circles indicate the positions of individual sensors arranged in 35 strings indicated by vertical gray lines. The true starting position and direction of the track are shown by the black line and arrow.
    The displayed event is a muon neutrino with an energy of 100~GeV that undergoes a charged-current interaction, producing a shower of particles at the interaction vertex and a long-ranging charged muon track.  In this particular interaction, about 2/3 of the neutrino's energy was deposited in the shower, resulting in a roughly isotropic ball of light at the vertex position, and the remaining 1/3 was given to the outgoing track.
    The colored circles represent sensor hits, with the color denoting the relative time of the earliest hit in a given module, and the size of each circle proportional to the amount of light it received. 
    }
    \label{fig:event_display}
\end{figure}

An event consists of a collection of hits. They can come from multiple sensors, and a single sensor can also contribute more than one hit. \Cref{fig:event_display} shows an example event display.
With regard to machine learning applications, it is important to note that the number of hits as well as where they appear in the detector can vary strongly between different events. A successful machine learning method must therefore accept a wide range of possible input data.

\subsection{Extended maximum likelihood}
\label{subsec:EML}

The technique of \ac{EML} is well suited to such inference tasks~\cite{Barlow:1990vc}. 
Per sensor, individual \ac{EML} likelihoods $\mathcal{L}_s$ can be formulated containing a Poisson ($\mathcal{P}$) term for the total number of observations per sensor (\eg, charge) and the probability $p$ of individual time(s) of observations (\eg, hits). The joint likelihood over all sensors of a detector is then simply the product over the \ac{EML} terms of all sensors, resulting in:
\begin{equation}
    \label{eq:dom_formulation}
    \mathcal{L}(\params|\obs) = \prod_{s=1}^{N_{\mathrm{sens}}} \mathcal{L}_s(\params|\obs) = \prod_{s=1}^{N_{\mathrm{sens}}}\left[\prod_{i=1}^{N_s} p_s(x_{i, s} | \params) \right] \mathcal{P}(N_s | \Lambda_s(\params)),
\end{equation}
where $s$ is the sensor index, $N_{\mathrm{sens}}$ is the total number of sensors, and $\Lambda_s(\params)$ the expected number of hits in sensor $s$ given $\params$. All quantities have $s$ subscripts because the hit \ac{PDF}, expected number of hits, and observed number of hits vary between sensors. The above treatment assumes that the statistical variations for each sensor are independent, which is typically the case. Due to the explicit product of individual terms of each sensor, we call this the ``\perdom'' formulation.

A mathematically equivalent formulation can be achieved by rearranging some of the terms in \cref{eq:dom_formulation}, ridding the explicit product:

\begin{equation}
    \label{eq:total_formulation}
   \mathcal{L}(\params|\obs)  = \left[\prod_{i=1}^{N_{\mathrm{tot}}} q_{s_i}(x_i | \params) \right] \mathcal{P}(N_{\mathrm{tot}} | \Lambda_\mathrm{tot}(\params)).
\end{equation}
where only one Poisson term for the total number of observed hits, $N_{\mathrm{tot}} = \sum_s^{N_{\mathrm{sens}}}N_s$, enters, and the product runs over all hits in the full detector with the new quantity  $q_s(x | \params) \coloneqq \frac{\Lambda_s(\params)}{\Lambda_\mathrm{tot}(\params)} p_s(x | \params)$. Consequently, $\Lambda_\mathrm{tot}(\params)$ is the expected number of hits in the entire detector.
This is still in the form of an \ac{EML}, but in this formulation $s$ has taken the role of an observable quantity rather than a label. Indeed, $q_s(x | \params)$ is a joint probability over $x$ and $s$. We will refer to this as the ``\alldom'' formulation.

While the expressions in \cref{eq:dom_formulation} and \cref{eq:total_formulation} are mathematically equivalent, the \alldom formulation is less computationally intensive.  The \perdom formulation (\cref{eq:dom_formulation}) features two explicit products, one over all sensors and one over all hits, while the \alldom formulation (\cref{eq:total_formulation}) involves just one product over all hits depending on $q_s(x | \params)$ and a single Poisson term that depends on $\Lambda_\mathrm{tot}(\params)$. 
A derivation of Eq.~\ref{eq:total_formulation} is provided in \cref{sec:EML_appendix}

\subsection{Toy Experiment}
\label{subsec:toy_experiment}

\begin{figure}
    \centering
    \includegraphics[width=0.7\textwidth]{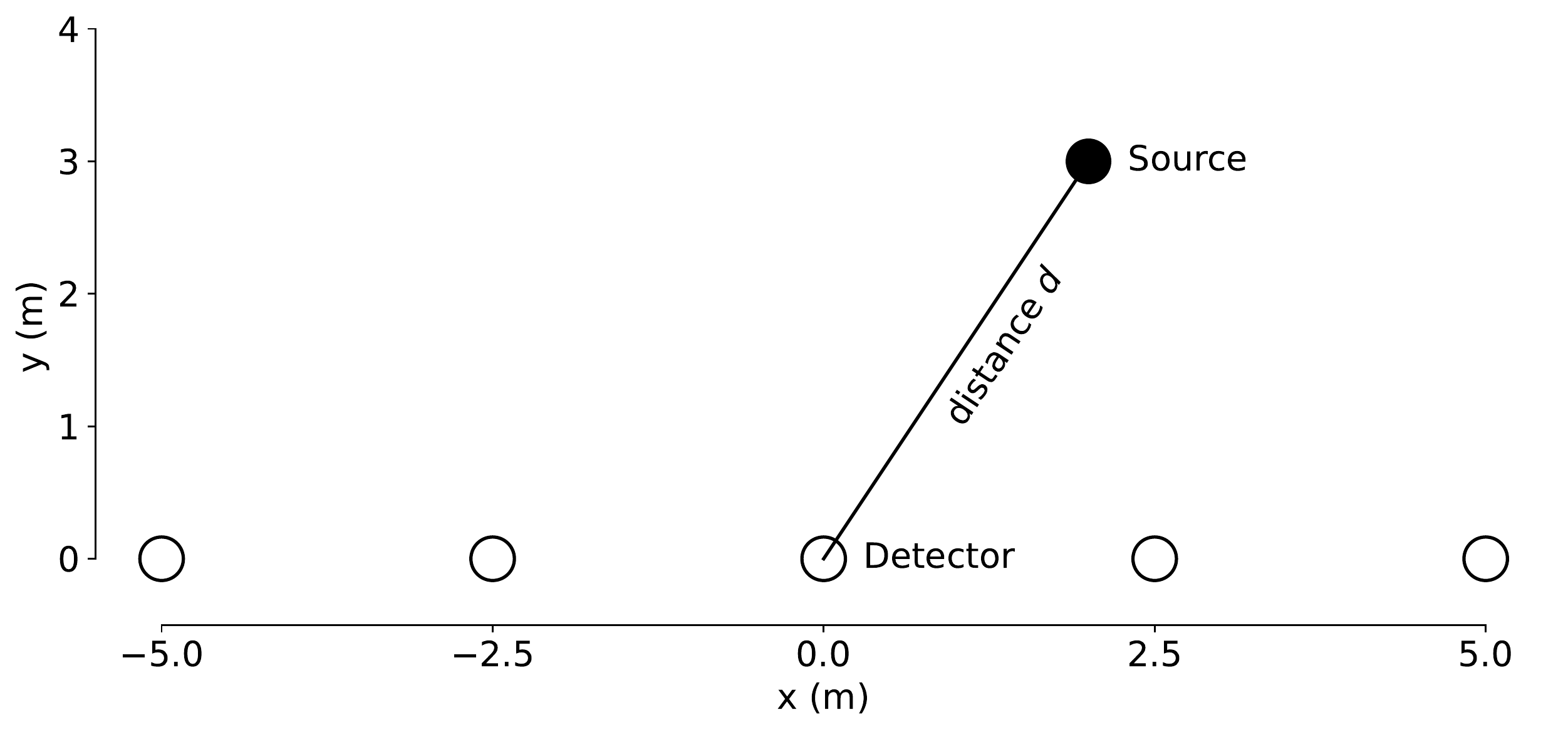}
    \caption{Illustration of our toy experiment, similar to Gull's lighthouse problem. A source at a given position \params (here $x_{src}, y_{src} = [2,3]$) emits light, and five detectors in a line register any incoming photons---the observations \obs.}
    \label{fig:gull}
\end{figure}

\begin{figure}[htbp]
\centering
\includegraphics[width=0.8\columnwidth]{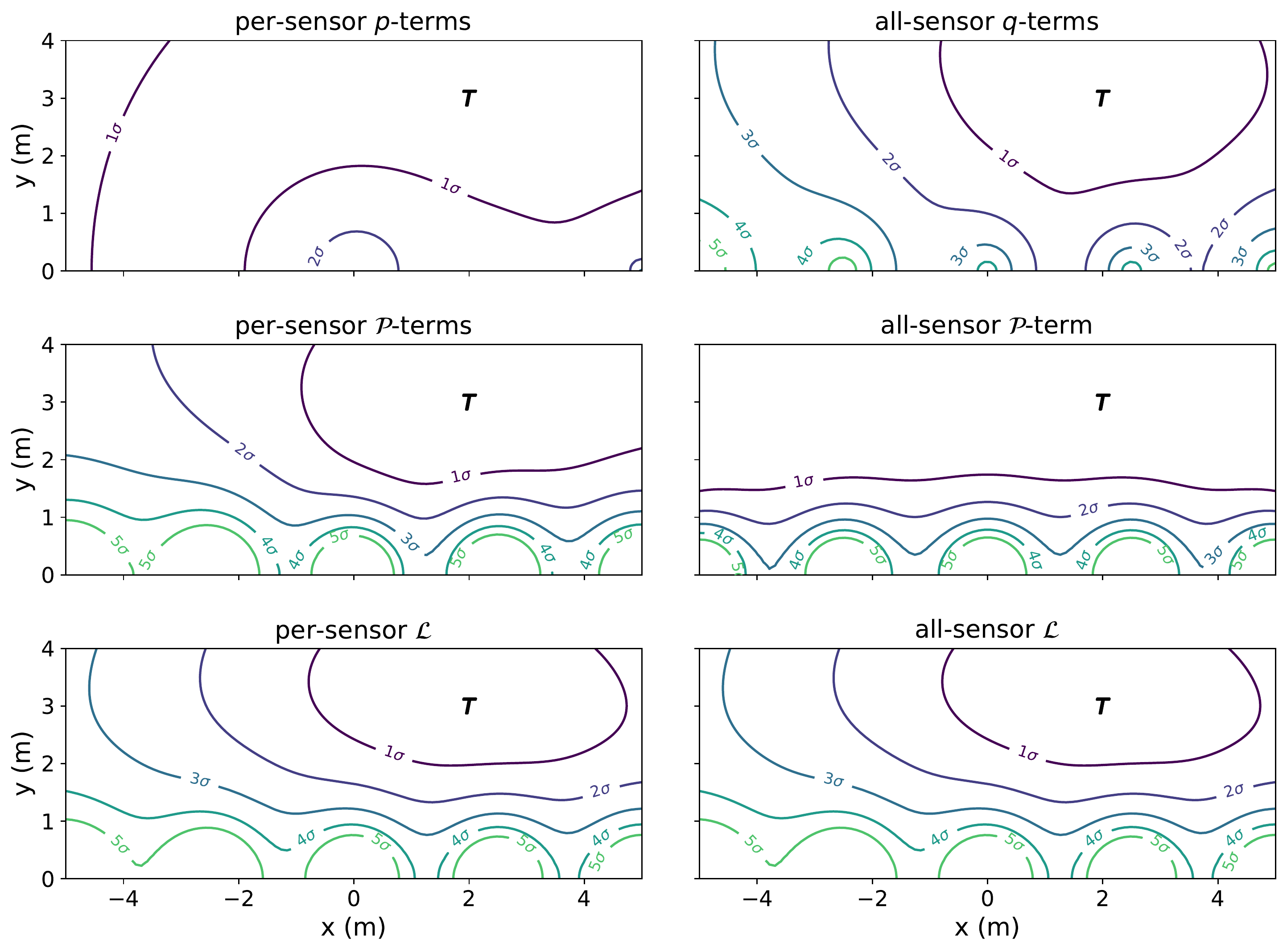}
\caption{Two formulations of the extended likelihood for an event from the illustrative toy example described in the text. Here we show the asymptotic 2-d confidence level (CL) contours for five standard deviations. The first column of plots depicts terms from the \perdom formulation (\cref{eq:dom_formulation}, and the second column depicts terms from the \alldom formulation (\cref{eq:total_formulation}). The ``\textit{T}'' marker indicates the true location of the source. The two formulations yield the same total likelihood (last row), but the \alldom formulation does not require an explicit sum over all sensors.}
\label{fig:eml_decomp}
\end{figure}

We will use a toy experiment to illustrate the difference between the two extended likelihood formulations. The experiment is similar to Gull's lighthouse problem~\cite{Gull1989}, but differs in the sense that we use a finite, discrete set of detectors and include timing information.
The setup is illustrated in~\cref{fig:gull} and features five photosensors arrayed in a line and a source that emits an instantaneous pulse of light at $t=0$. Each sensor then reports a number of pulses drawn from a Poisson distribution whose mean decreases as a function of the distance $d$ from the light source to the sensor. Each individual pulse occurs at a time drawn from an arrival time distribution that is also a function of the distance $d$ from the sensor to the source. More details on the specific functional forms can be found in \cref{sec:toy_appendix}.
In this example, the hypothesis \params describes the position of the light source and has two parameters: $x_{\mathrm{src}}$ and $y_{\mathrm{src}}$. \Cref{fig:eml_decomp} shows likelihood scans for this toy experiment, as well as the $p$ and $q$ terms from \cref{eq:dom_formulation} and \cref{eq:total_formulation}. The two formulations conceptualize and decompose the likelihood function differently but ultimately yield the same result.

\section{Likelihood Ratio Estimators}

The likelihood function needed for reconstruction, such as those in \cref{eq:dom_formulation} or \cref{eq:total_formulation}, describes underlying complicated processes such as particle interactions, photon scattering, and the physical and electronic responses of the detector hardware specific to each experiment. Generally speaking, accounting for all these effects renders the exact likelihood intractable, and some degree of approximation is required.
At the same time, it is common for particle physics collaborations to develop sophisticated forward Monte Carlo simulations that enable them to generate samples from $p(\obs | \params)$ for any $\params$. There are many ways to leverage these simulations to approximate the functions needed to evaluate \cref{eq:dom_formulation}; here we present one such approach that works for arbitrary detector geometries, scales well to detectors with many discrete sensors, and minimizes the amount of assumptions and approximations by requiring none beyond those already used in the detector simulation.

Using a forward $ \params\rightarrow\obs $ simulation, one can train an \ac{ML} model $\mathrm{d}(\obs, \params)$ to classify samples as being drawn from $p(\obs, \params)$ (labeled $y=1$) or $p(\obs)p(\params)$ (labeled $y=0$). The $y=1$ samples are unmodified simulation events, and the $y=0$ samples are produced by drawing \obs and \params independently from the simulation data set (\ie, by choosing \obs from one randomly selected event and \params from another randomly selected event). When trained with binary cross-entropy loss, the optimal classifier $\mathrm{d}^*(\obs, \params)$ will satisfy Eq.~\ref{eq:optimalclassifier}~\cite{hermans2020likelihoodfree}:
\begin{equation}
    \label{eq:optimalclassifier}
    r(\obs, \params) \mathop: = \frac{\mathrm{d}^*(\obs, \params)}{1 - \mathrm{d}^*(\obs, \params)} = \frac{p(\obs | \params)}{p(\obs)}.
\end{equation}
Hence, such a classifier approximates the likelihood-to-evidence ratio $r$. For reconstruction tasks, the likelihood-to-evidence ratio can replace the likelihood function because the scaling factor $p(\obs)$ is independent of \params, \ie, constant. Therefore, a classifier trained to approximate this ratio---a ratio estimator---is sufficient to conduct parameter inference.

Unfortunately, training a single classifier $\mathrm{d}(\obs, \params)$ is not straightforward when \obs is a variable-length series of hits. There are techniques for representing variable-length input in a format suitable for a machine learning model such as a neural network (e.g. \cite{Abbasi:2021ryj, KM3NeT:2020zod, Baldi:2018qhe}), but these techniques often result in information loss, assume certain detector properties, or require complicated model structures.
Some approaches that have been proposed and used to deal with irregular involve, for example, the usage of interaction networks \cite{battaglia2016interaction} or graph neural networks \cite{IceCube:2022njh}.
Rather, we propose to train classifiers for each of the likelihood terms in \cref{eq:dom_formulation} or \cref{eq:total_formulation}, respectively. The derived ratio estimators then replace the probability distributions when evaluating the extended likelihood. This approach allows for simpler model architectures and can be easily adapted to a wide variety of different experiments and detector configurations. 

\begin{figure}[htbp]
\centering
\includegraphics[width=0.8\columnwidth]{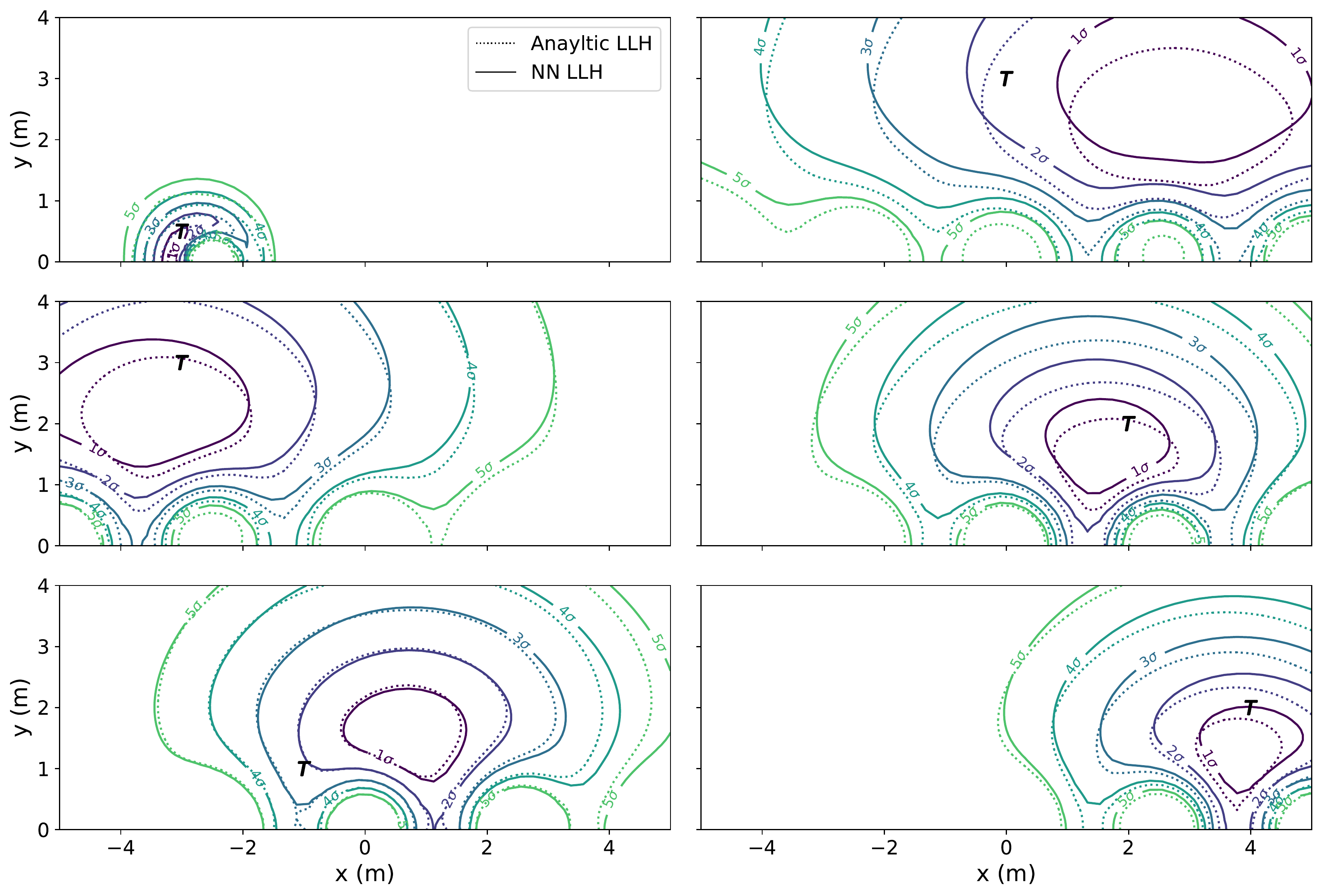}
\caption{Approximation of the extended likelihood function for the 2d illustration example for six example events. The true \ac{LLH} is shown in the dotted CL contours, while the \ac{NN} approximation is overlaid in the solid lines. The \acp{NN} are trained as a ratio estimators as described in the text and captures the salient features of the true likelihood function. The ``\textit{T}'' marker indicates the locations of the source for each event.}
\label{fig:example_toy_nns}
\end{figure}

\cref{fig:example_toy_nns} shows the result of applying this technique to the toy experiment described in \cref{subsec:toy_experiment}. We replace the time \ac{PDF} and Poisson term with two \ac{NN} ratio estimators: $\hat{r}_{\mathrm{hit}}(t, s, \params)$ replaces $p^\mathrm{tot}_{s}(\obs | \params)$, and $\hat{r}_\mathrm{tot}(N_{\mathrm{tot}}, \params)$ replaces $P(N_\mathrm{tot} | \params)$. As in \cref{subsec:toy_experiment}, \params represents the hypothesized true position of the light source and has two components: $x_\mathrm{src}$ and $y_\mathrm{src}$ positions of the source. The resulting likelihood landscapes (analogous to those shown in \cref{fig:eml_decomp}) match their true counterparts well, especially near the maximum of the likelihood function.

\subsection{Implementation}
In principle, any binary classifier can be used for the technique described above. Following~\cite{hermans2020likelihoodfree}, we use fully connected artificial \acp{NN}.
These networks cannot handle a variable-length input, such as a variable-length collection of hits. To solve that issue, we feed the individual hit information, time and position of a single hit, one after another to a network (called \hitnet) and sum up the individual outputs to get the joint log-likelihood (\ac{LLH}).
In this way, the network always sees the same input structure, but we lose information about sensors without hits, \ie, the information about the total number of expected hits. To get it back we train a second fully connected network (called \chargenet) to predict the number of expected hits and also include its output to obtain the complete likelihood.
In our examples, we assume sensors in a detector to be identical and hence train a single \hitnet. If sensors of different types are used, multiple {\hitnet}s can be trained.

As described in~\cref{subsec:EML},~\cref{eq:dom_formulation,eq:total_formulation} give two different ways of splitting up the \ac{EML} into its component individual hit probability and expected total number of hits. 
The training process differs between the different versions of the networks. In general, the training process introduced in~\cite{hermans2020likelihoodfree} requires showing the networks correlated combinations of \obs and \params (labeled ``1'') and uncorrelated combinations (labeled ``0''). What changes for the different versions of the networks is the observation \obs in case of the {\chargenet}s and the way to shuffle \obs and \params for the {\hitnet}s.

\paragraph{Approximation of the \perdom Terms}
The \chargenet gets the total number of hits in a single sensor as input for a specific event and approximates $\mathcal{P}(N_{s} | \params)$ as shown in~\cref{eq:dom_formulation}. As with the \hitnet, there are multiple observations per event, one per sensor, that have to be combined to get the likelihood. For the \hitnet, \obs is shuffled in-sensor, meaning that only observations from the same sensor are used for the shuffling. The network thus learns $p_{s}(\obs | \params)$.

\paragraph{Approximation of the \alldom Terms}
The \chargenet gets the total number of hits in the detector as input to approximate $\mathcal{P}(N_\mathrm{tot} | \params)$ as shown in \cref{eq:total_formulation}. For the \hitnet, \obs is shuffled between all sensors, allowing the network to learn $q_{s}(\obs | \params)$.\\

\begin{table}[]
\centering
\includegraphics[width=0.5\textwidth]{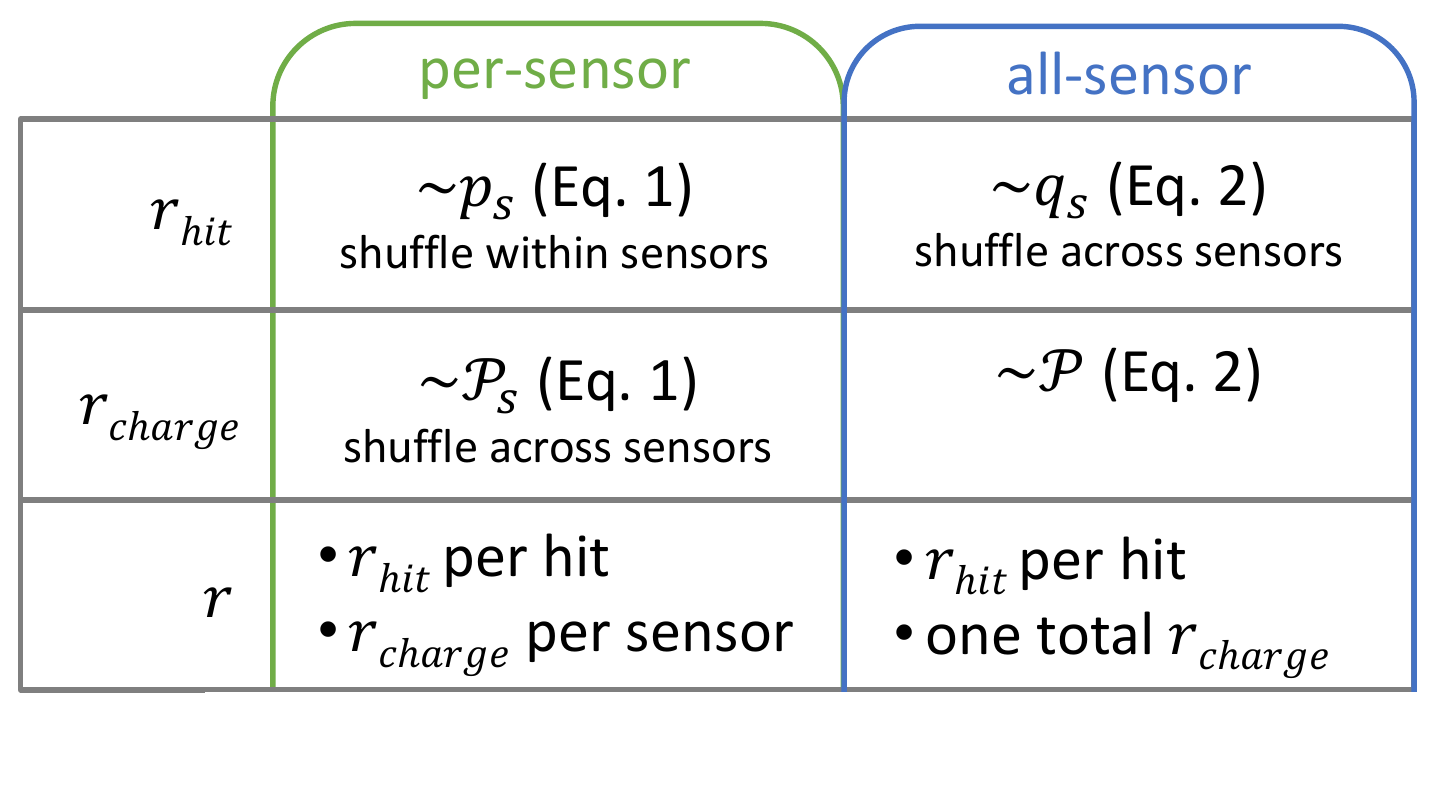}
\caption{Overview of the different formulations (\perdom and \alldom) and their necessary ratio estimators to build the full ratio estimator $r$ proportional to the likelihood. The shuffling strategy for the uncorrelated sets in the training is summarized for the separate terms.}
\label{tab:my-table}
\end{table}

\begin{figure}[htbp]
\centering
\includegraphics[width=0.9\columnwidth]{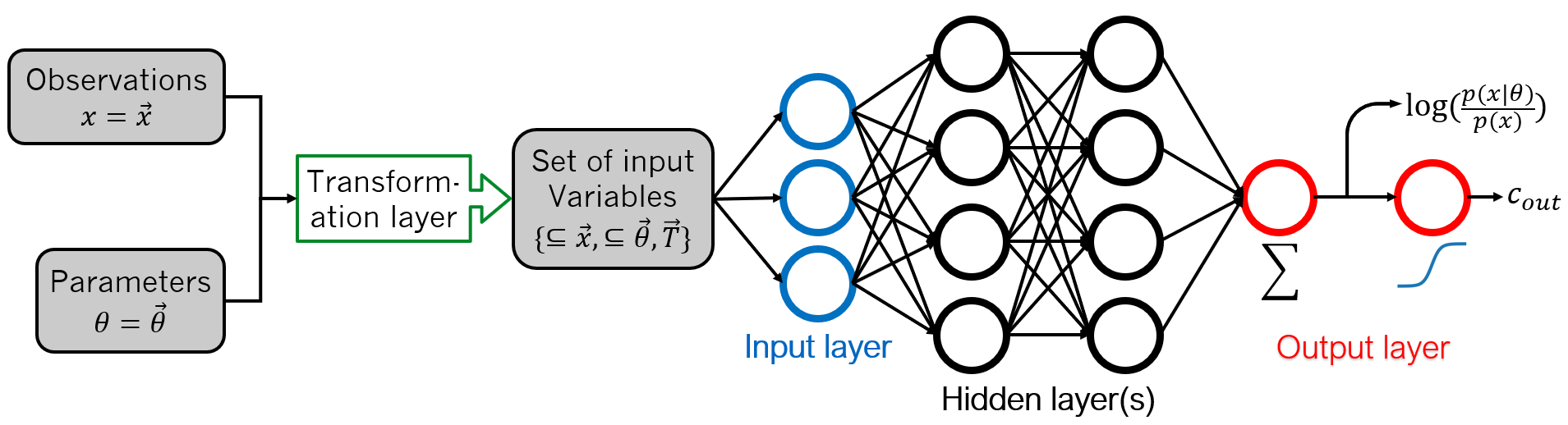}
\caption{Input processing structure used for likelihood-free inference, including a transformation layer and a neural network. $\vec{T}$ represents variables calculated from $ \obs $ and $ \params $, e.g. distance between sensor and light source. $ c_{out}~\epsilon~[0,1] $ is the binary classifier output.}
\label{fig:nn_structure}
\end{figure}

These are the essential procedures to learn the likelihood. In addition, to further improve the network performance, we used a transformation layer as first network layer. Quantities the likelihood should depend on, for example, the relative distance between sensor and interaction vertex, can be calculated in that layer and passed to the network instead of directly passing \obs and \params. It is also advantageous to standardize input distributions such that all inputs span over similar ranges for better and faster training convergence; the necessary transformations can also be performed in this first layer.
In contrast to the activation in the last layer, the activation functions in all other layers can be chosen freely. We use a smooth activation function (\eg, \textsc{swish} \cite{ramachandran2017searching}), otherwise the learned likelihood will not be smooth, which is particularly important for evaluating gradients of the likelihood.
To directly obtain the logarithm of the likelihood-to-evidence ratio from the networks, the activation function in the last layer can be changed post-training from sigmoid to linear.\\
Figure~\ref{fig:nn_structure} illustrates the structure used for likelihood-free inference learning of the likelihood-to-evidence ratio.

\section{Example Applications}
In this section we present a few example applications for the technique discussed above. All examples are based on the toy experiment described in~\cref{sec:toy_appendix}, and an open source implementation for reproducibility is available\footnote{Software located at \url{https://github.com/philippeller/freeDOM}, the first example (Sec.~\ref{sec:spherical_detector}) can for instance be reproduced using the notebook under \texttt{notebooks/spherical\_toy\_paper.ipynb}}. A toy model is advantageous because one can readily access the true likelihood for comparison purposes. Applications to a realistic detector will be discussed in \cref{sec:tonscale}.
Different detector configurations (sensor positions) are used in each example, highlighting the intrinsic flexibility of the technique.
The following eight parameters (\ie \params) are used in our event model:

\begin{itemize}
\item \textbf{Position} $\bm{(x, y, z)}$: Three dimensional position of the interaction in the detector.
\item \textbf{Time} $\bm{t}$: Time of the interaction relative to the first trigger time.
\item \textbf{Direction angle}s $\bm{(\varphi, \vartheta)}$: Azimuth and zenith angle of the direction of the incoming neutrino.
\item \textbf{Total deposited energy} $\bm{E}$: Energy of all non-neutrino particles produced after the neutrino interaction.
\item \textbf{Inelasticity} $\bm{I}$: Fraction of the total energy that goes into the cascade.
\end{itemize}

\subsection{Validation: Toy Spherical Detector}
\label{sec:spherical_detector}
We first consider a toy detector consisting of a 162 sensors arrangement on a sphere with a radius of 10~m. For the positioning of the individual sensors we used the \textit{icosphere} python package~\cite{icosphere} that generates points uniformly distributed over the sphere.
The training set and the test set used for reconstruction consist of events that are completely contained in the detector, \ie, the interaction vertex and the last track element are inside the detector volume.
Four networks were trained for this toy detector: one \hitnet and one \chargenet for each of the \alldom and the \perdom formulations.

Since our likelihood is a function of eight parameters, it is a challenge to quantify the agreement with truth or even visualize the function outputs. To verify the quality of the trained networks, we follow a suggestion from \cite{hermans2020likelihoodfree}. The idea is to reweight the distribution $p(\obs)$ of measurement variables over the full training set by our ratio estimator $\hat{r}(\obs,\params_{i})$ to obtain the conditional distribution $p(\obs|\params_{i})$ for a specific set of model parameters $\params_{i}$. Such reweighted distributions can then be compared to the respective analytic conditionals.

\begin{figure}[h]
    \centering
    \includegraphics[width=\textwidth]{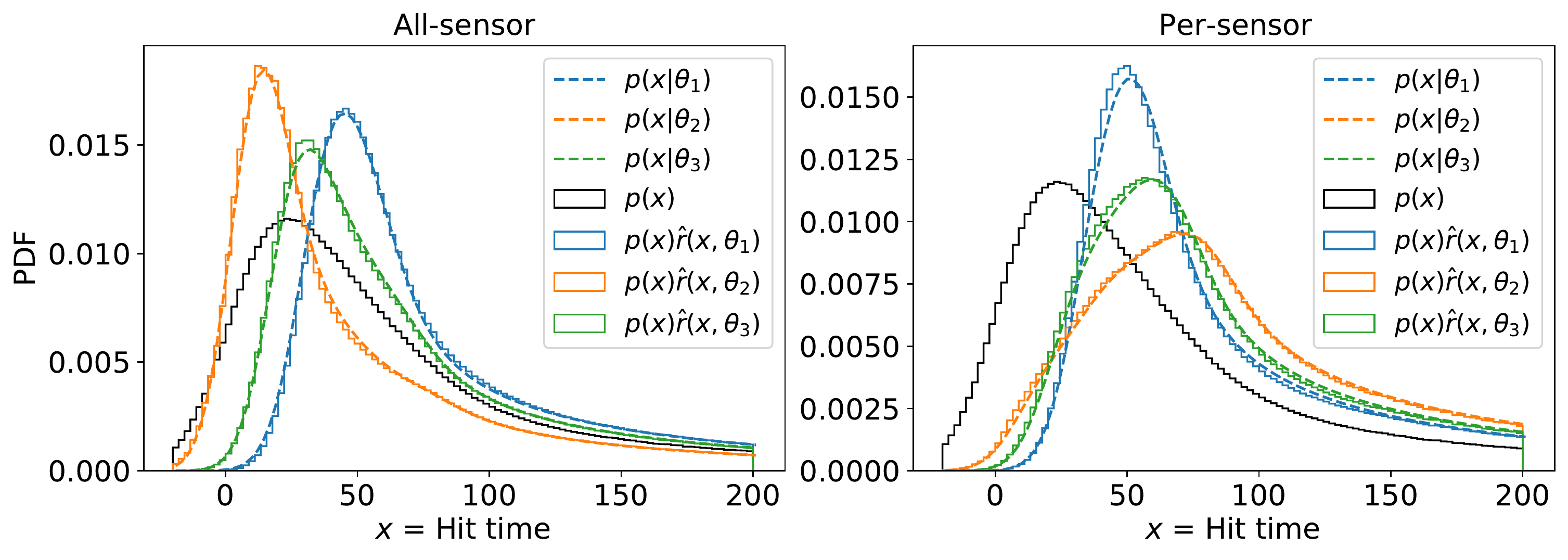}
    \caption[Spherical toy reweighted hit time distributions]{Likelihood verification based on event reweighting. The black distributions show the times of all hits in the sample the network was trained on. The dashed lines represent the hit time PDFs for three example events. The colored distributions are obtained by reweighting the black distributions according to the respective event parameter values. In the left plot the \alldom formulation is used and in the right the \perdom formulation.}
    \label{fig:spherical_reweight_time}
\end{figure}

Our observables \obs are the time and position of a hit (mainly used by the \hitnet) as well as the charge of the hit (used by the \chargenet). 
This means we can assess the \hitnet performance by looking at the time distributions of the hits. \Cref{fig:spherical_reweight_time} shows the times of all hits over all events in a sample similar to the training sample (black). This is $p(\obs)$ for the time component. We show conditionals for three different sets of event parameters from our analytic truth, together with the reweighted $p(\obs)$ distribution for each of the three sets.
In the left panel, an \alldom \hitnet was used for the reweighting and in the right panel, a \perdom \hitnet. The difference between these comes from the fact that they describe different parts of the decomposed likelihood and consequently have distinct PDFs. In the \alldom formulation, the \chargenet does not contribute individual sensor information, so the \hitnet has to learn a time distribution (from~\cite{vanEijndhoven:2007ge}) multiplied by the probability for the photon to reach the respective module. In contrast, in the \perdom formulation the \chargenet provides information about individual sensors, and the probability for a photon to reach a sensor is part of the charge PDF. As a consequence, the \hitnet only has to learn the arrival time distribution. That is the reason why it is trained by only shuffling information within a sensor. All reweighted distributions agree well with the analytic truths, indicating that the networks have approximated the likelihood well at these points in the \params space.

\begin{figure}[h]
    \centering
\includegraphics[width=\textwidth]{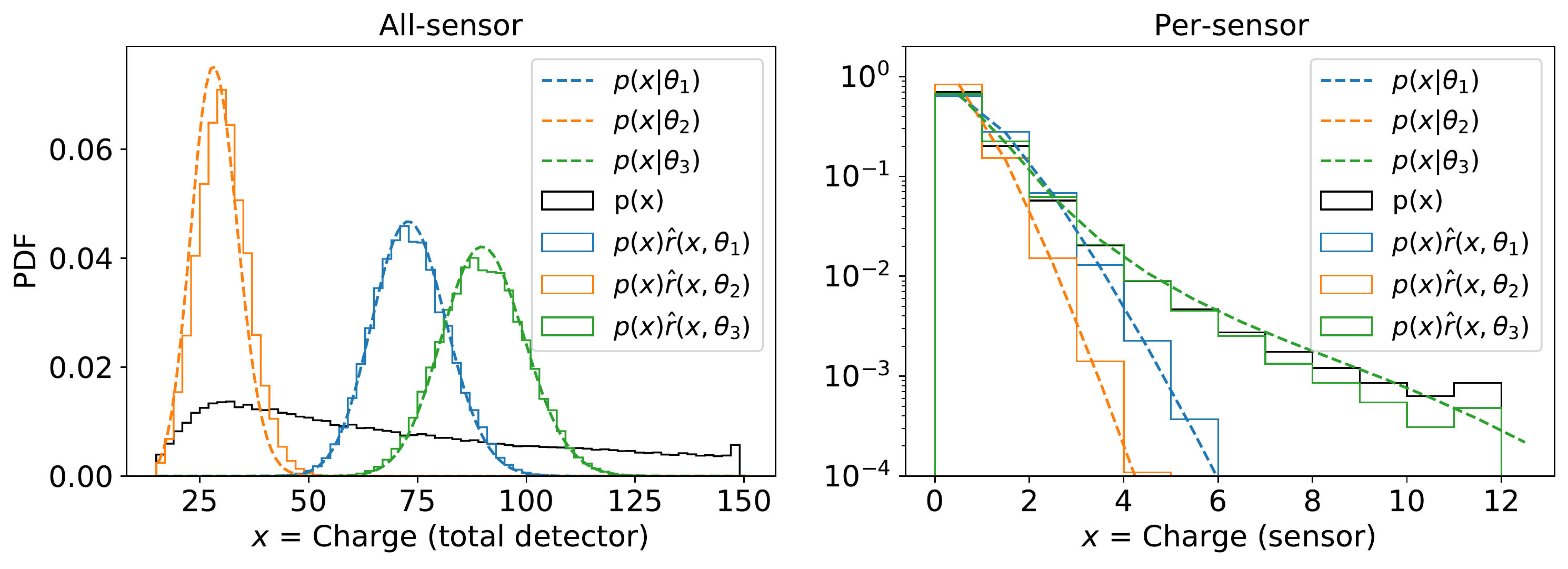}
    \caption[Spherical toy reweighted charge distributions]{Likelihood verification based on event reweighting. In the left plot, the black distribution shows the charge deposited in the complete detector for all events in the training sample. The right plot shows the charges for (all) individual sensors. The dashed lines represent the respective charge PDFs for three example events. The colored distributions are obtained by reweighting the black distributions according to the respective event parameter values.}
    \label{fig:spherical_reweight_charge}
\end{figure}

The same reweighting can be done for the charge of a hit using the \chargenet. \Cref{fig:spherical_reweight_charge} shows the total detector charge (left) and per-sensor charge (right) distributions. Again, the distributions for the event sample similar to the training sample are shown in black. The reweighted distributions for three example sets of event parameters are shown as solid lines, overlaid with the theoretical expectations as dashed lines. The \chargenet approximates the ratio $r$ well for the tested event parameter values.

\begin{figure}[h]
    \centering
    \includegraphics[width=\textwidth]{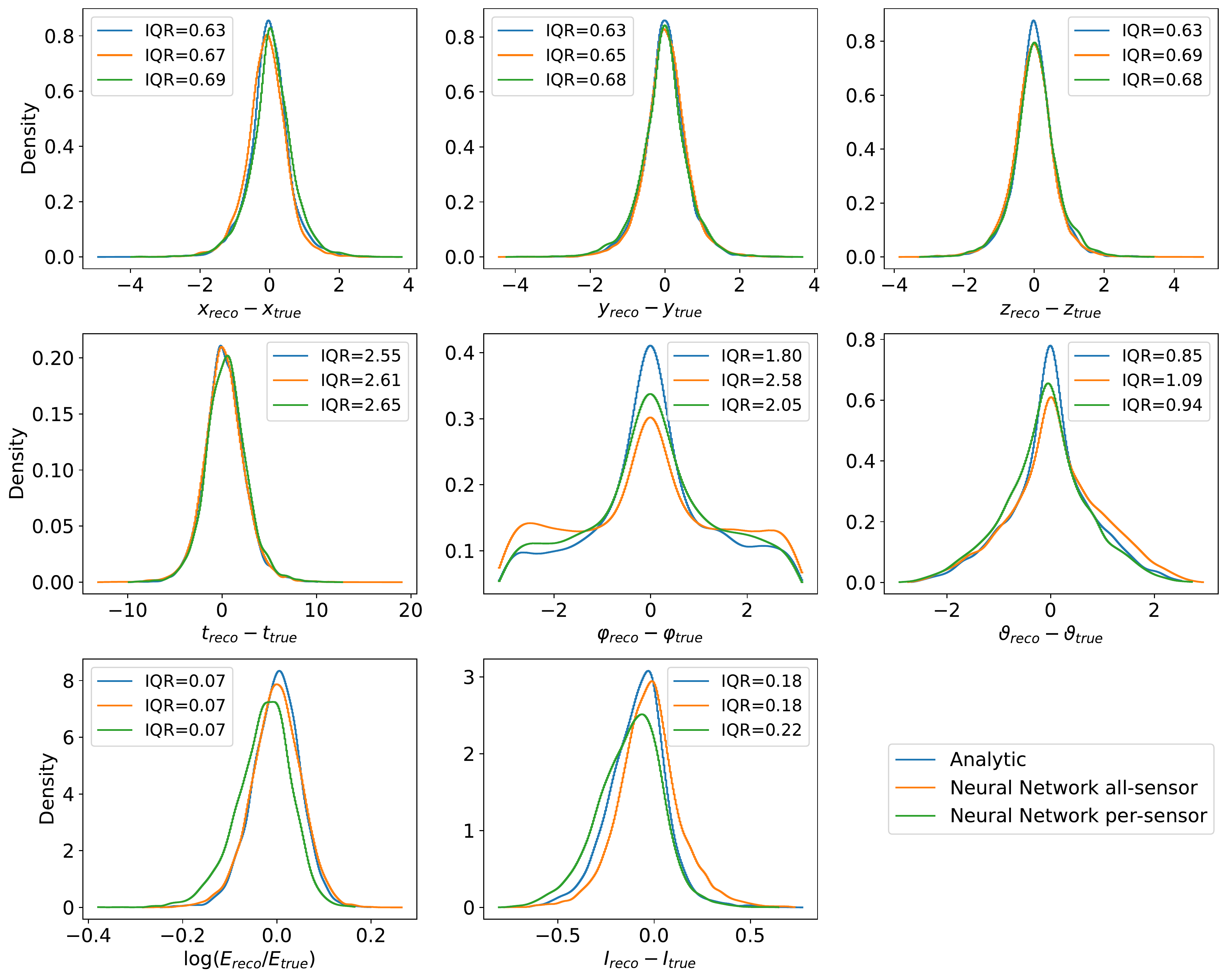}
    \caption[Spherical toy parameter resolutions]{Parameter resolutions of networks trained in both likelihood formulations compared to the resolutions obtained with the true analytic likelihood (which is the same in both formulations). The distributions are plotted as KDEs.}
    \label{fig:spherical_toy_res}
\end{figure}

The two tests illustrated in Figs.~\ref{fig:spherical_reweight_time}~and~\ref{fig:spherical_reweight_charge} show the fidelity of the techniques for a subset of three points in the event parameter space.
To go a step further, we perform a reconstruction of events. We perform this reconstruction by finding the maximum likelihood estimators for \params for each event by running an optimizer on the (approximate) likelihood, and repeat this procedure for a large number of events sampled throughout our parameter space.
For a well-approximated likelihood (in the parameter space of interest), the resolutions obtained from a reconstruction using the learned function should match those using the analytic truth.
\Cref{fig:spherical_toy_res} shows the resolutions for both formulations compared to the resolutions using the true likelihood.

The networks in the two different formulations perform very similarly.
The resolutions obtained from the reconstruction with the true likelihood represent the theoretical limit and are slightly better than the network results, as expected. This means that the reconstructions with the neural network approximations are close to the limit, but there is still some potential for improvement. The limited number of training events as well as an imperfect training process are reasons for the difference.

While the reconstruction performance for the \hitnet and \chargenet formulations are very similar, there is a significant difference in computational time required. In the \perdom formulation, \chargenet has to be evaluated once per sensor to calculate the likelihood at a specific \params. However, in the \alldom formulation it has to be evaluated only once. 
Reconstruction with the \perdom networks takes about twice as long as with the \alldom networks. The speed difference depends roughly on the number of sensors in the detector minus the average number of hits in the event sample. This means that for detectors with large numbers of channels and sparse data, this difference can become much more dramatic.

\subsection{Detector Design: Sensor Density Optimization}
\label{sec:toy_detector_opt}

One of the big advantages of our technique is its flexibility. In the \perdom formulation the \hitnet and the \chargenet receive---as suggested by the name---only per-sensor information. If the transformation layer only passes relative and no absolute positional information, the networks become independent from the particular detector configuration. This means that networks trained on one specific detector configuration can be used to reconstruct any other configuration. This can be used to test the impact of different detector configurations on reconstruction resolutions without changing or retraining the networks.

\begin{figure}[h]
    \centering
    \includegraphics[width=0.7\textwidth]{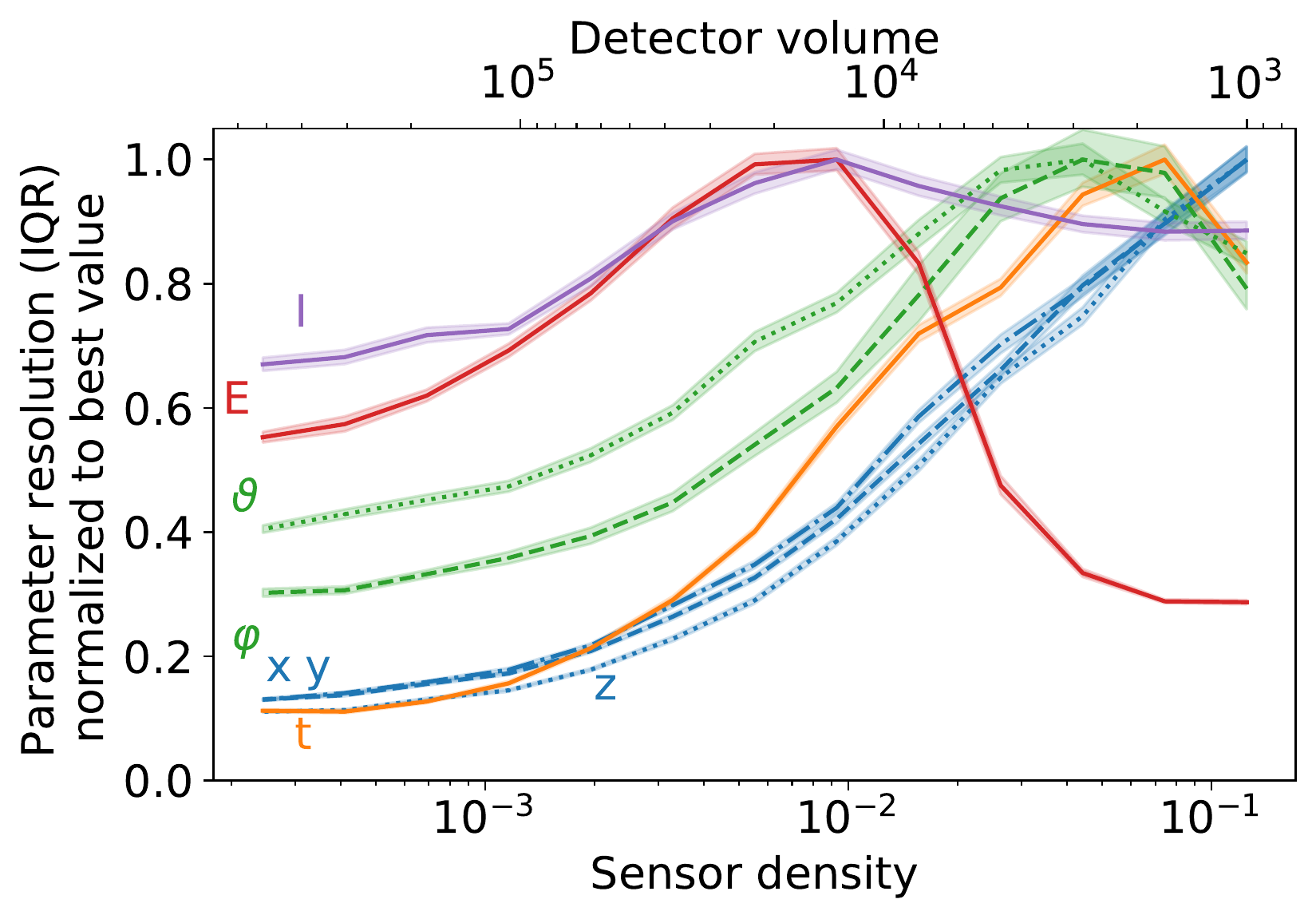}
    \caption[Detector optimization study]{The influence of sensor density on parameter resolutions for a fixed number of 125 sensors in a cubic geometry (with varying sensor distances). Parameter resolutions are quantified by the 50 percent inter quantile range. The relative resolutions compared to the best one are shown. The error ranges are determined via one-sample bootstrapping.}
    \label{fig:res_vs_dens}
\end{figure}

We use this property of our method to perform a detector optimization study using a toy model. A \hitnet and a \chargenet were trained on simulation with a single sensor. The events in the training sample were placed randomly in a sphere with a radius of 50\;m around the module to cover all possible sensor-event combinations. This simple detector configuration also makes it possible to simulate and train on a lot of events quickly.
The detector configuration then used for the reconstruction consists of a $5\times5\times5$ cubic grid of sensors. The distance between neighboring sensors is the same in all three dimensions. For the optimization study, the sensor density is varied without changing the number of sensors, only their distance to each other changes. The reconstructed event sample consists of tracks that are also generated in a cube of the size of the smallest detector configuration used here. Hence the generation volume is always completely contained in the detector, but the tracks are not necessarily so.
A minimum of four hits was required for each detector configuration, to only select reconstructable events. This has the effect that smaller detectors reconstruct more low-energy events, even though the events for all detectors were drawn from uniform distributions in all parameters.
\Cref{fig:res_vs_dens} shows the relative change in parameter resolution (central quantiles) for all eight parameters describing a neutrino interaction. The resolutions are shown as relative numbers with respect to the best performance at 1.0, meaning that, for example, 0.5 would indicate a 50\% decrease in resolution compared to the best performer among the tested detector configurations.

In general, the resolutions improve with more densely instrumented detectors. 
The three vertex variables show a virtually identical behavior and their resolutions improve with higher sensor densities. The correlated behavior is expected as the sensor is symmetric in all three spatial dimension, and also the improvements are as expected since a closer sensor spacing can better resolve positions.
The same is true for the reconstruction of the interaction time $t$.
For the remaining four parameters, the fact that longer tracks have a higher probability of not being completely contained in a smaller detector plays an important role. This can be seen most clearly in the energy resolution. If a track is not completely contained in the detector, there is light escaping, and the event seems to be less energetic, which also affects the angular resolutions.
So, for a given true energy range (physics signal region), a given number of sensors (detector cost), and a desired measurement goal, there is likely an optimal sensor density where the best overall resolution can be achieved. Our method can be used to identify this optimal value, thanks to its high flexibility. It is also possible to test completely different sensor arrangements to find the optimal one.
In our small example study, only the effect of sensor density on parameter resolutions was investigated. No additional effects of different effective volumes and trigger rates were considered.

\subsection{Bayesian Posterior via MCMC}
\label{sec:mcmc}
Having a function proportional to the likelihood means that it can also be used for Bayesian inference. For parameter inference, one needs to access the so-called posterior distribution, which encodes the information of model parameters after the Bayesian knowledge update with our observed data $x$.
Since the posterior distribution is, in general, not analytically available, one resorts to approximate techniques, such as sampling or variational inference. Here, we demonstrate the usage of our learned likelihood for MCMC sampling.
A converged MCMC chain will produce samples according to the posterior distribution, regardless of the normalization of the input density, which means it is irrelevant that our function is only proportional to the true likelihood, since it is divided by the evidence over the training set.

\begin{figure}[h]
    \centering
    \includegraphics[width=\textwidth]{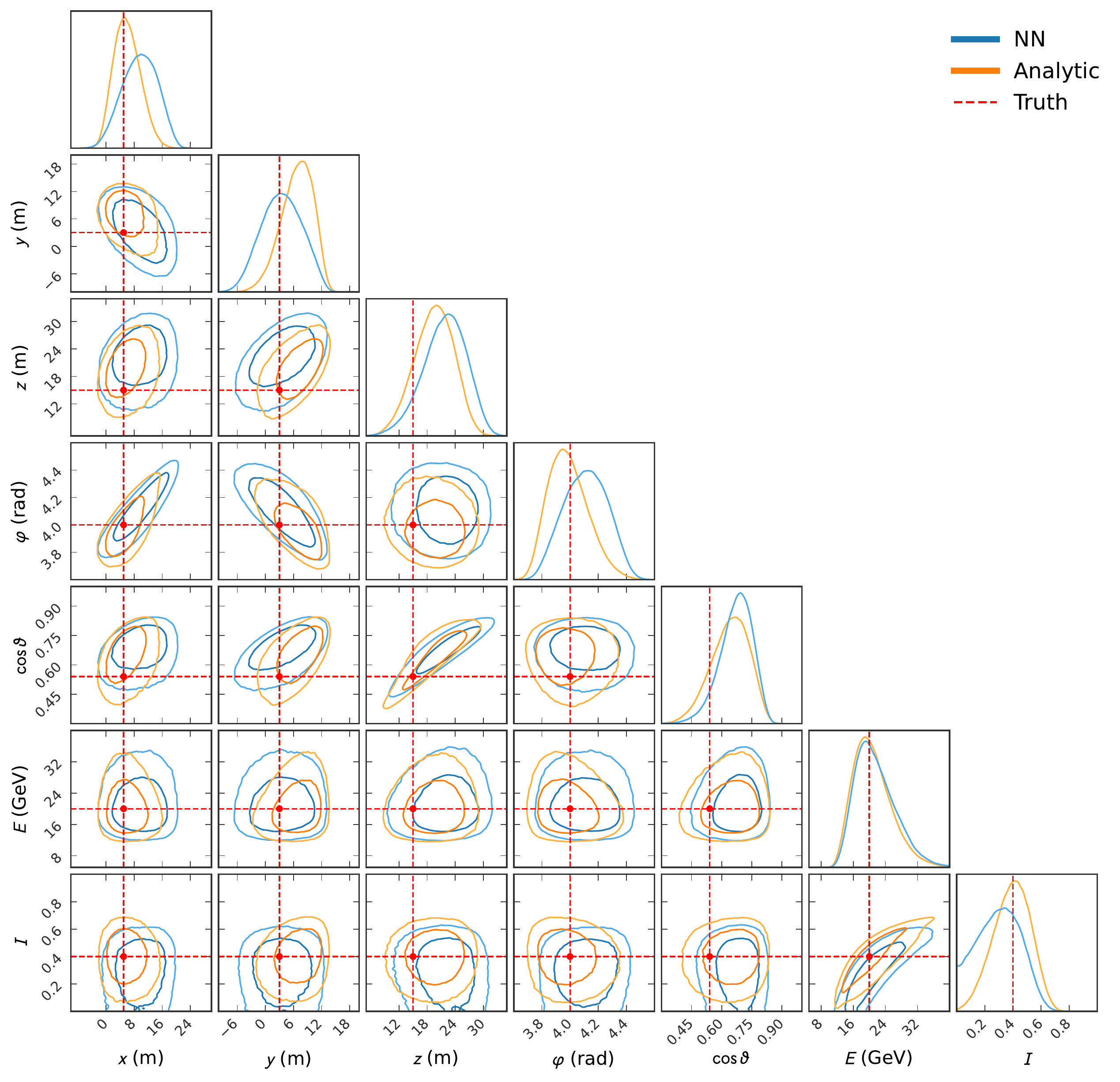}
    \caption{Marginal posterior distributions resulting from the MCMC sampling given the test event. The 1d (kernel density estimates) and 2d (68\% and 95\% credible regions) distributions are samples from our approximate likelihood (blue), and samples from the true likelihood (orange) for reference. The red dashed cross hairs and vertical lines indicate the true parameter values used to simulate the test event.}
    \label{fig:mcmc}
\end{figure}

For illustration, we continue using the same toy experiment setup and construct a neutrino telescope with several strings arranged in a sunflower pattern in the $x$--$y$ plane.
A rendering of the detector together with an example event is shown in \cref{fig:event_display}.
We have trained a model in the \alldom formulation on $10^6$ simulated events, and use the resulting surrogate likelihood function with a Metropolis-Hastings sampler~\cite{Schulz:2021BAT} to draw $10^5$ samples from the 8-dimensional posterior distribution.
For comparison and validation purposes, we can also use our true likelihood function to generate MCMC samples from the true posterior distribution with the same sampler.
\Cref{fig:mcmc} shows the resulting posteriors marginalized to  1d and 2d, together with the true parameter values known from the simulation of a test event. The different marginalized posteriors in the figure demonstrate good agreement between using our approximation and the true likelihood, whereas the former results in slightly wider distributions. This widening is the same effect as discussed in \cref{sec:spherical_detector} and can be attributed to a classifier slightly less performant than the theoretical optimum.
The correlations between different parameters are captured well, as can be seen from the two dimensional marginal distributions, such as $\cos{\theta_\mathrm{zen}}$ (coszen) vs. $z$ position, azimuth vs. $x$ and $y$ positions, or inelasticity vs. energy.

Since our likelihood is based on neural networks, it means that gradient information is also available. This opens the door to use more advanced sampling techniques, such as Hamiltonian MC \cite{Betancourt2017ACI}, as well.

\section{Hybrid Neutrino Detector}
\label{sec:tonscale}

In this section, we apply the preceding inference technique to a cylindrical ton-scale neutrino detector.  The photosensor layout varies along the detector geometry, with the photocathode coverage decreasing as a function of detector height.  The asymmetric coverage is chosen since it has a significant impact on the detector response as a function of position and, in particular, direction.  This detector is filled with a medium comprised of water and scintillator mixture, called Water-based Liquid Scintillator (WbLS)~\cite{Yeh:2011}.  In WbLS, both directional Cherenkov light and the slightly delayed, isotropic scintillation light are produced when a neutrino interacts within, in principle providing pointing at lower energy thresholds than other media.

To achieve a successful separation of the Cherenkov and scintillation light signals, there are a number of geometrical, light collection, and detection medium properties that can be optimized.  Obtaining a quantitative measure of detector performance while varying these parameters is a crucial step in the design of a realistic detector.  Compared to the previous toy models in this paper, this detector study required a more precise Monte Carlo simulation to capture the intricacies of particle dynamics and detector responses.  

RAT-PAC~\cite{ratpac}, a Monte Carlo simulation toolkit based on Geant4~\cite{GEANT4:2002zbu}, was employed to accomplish this task.  This simulation provides a more realistic model for the detector by implementing complex photon dynamics from production to capture and the corresponding photosensor response.  The details of the MC model for the WbLS can be found in~\cite{Land:2020oiz}.  The light yield, time profile and emission spectrum of scintillation light are based on measurements of WbLS mixtures that have 1, 5 and 10\% scintillator fraction by mass~\cite{D0MA00055H,Caravaca:2020lfs}, and these results are interpolated and extrapolated in order to predict the properties of the particular mixtures considered here.  A model for scattering and attenuation is assumed, which is described in~\cite{Land:2020oiz}.  We simulated large datasets for each detector configuration, and used the correlated and derived uncorrelated (\params,\obs) to obtain a surrogate likelihood that could then be used to reconstruct and study fresh data.
A more traditional approach to this type of study, using likelihood-based reconstruction and a similar set of software tools, is described in Ref.~\cite{Land:2020oiz}.

\begin{figure}[h]
    \centering
    \includegraphics[width=0.6\textwidth]{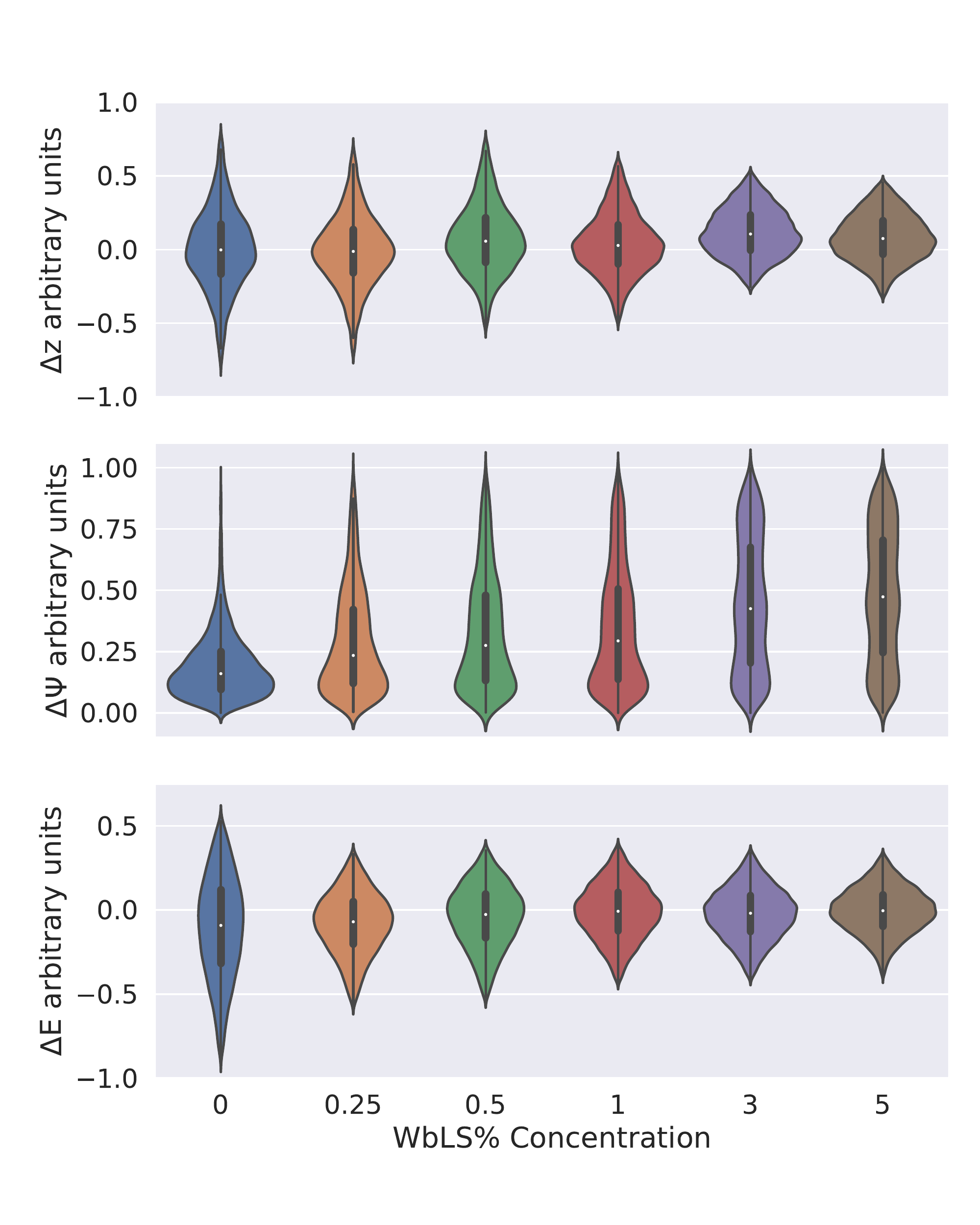}
    \caption{A study of resolution performance on a preliminary design configuration for a ton-scale detector as a function of WbLS concentration.  Here, $\Delta z$ is the difference between reconstructed and true particle height, $\Delta \Psi$ is the angle between the true and reconstructed particle direction, and $\Delta E$ is the difference between reconstructed and true particle energy. All units are relative arbitrary units (a.u.) As the scintillation light yield increases, the vertex and energy resolution show a marked improvement, while the directional resolution degrades. }
    \label{fig:tonscale_reco}
\end{figure}

Here, our deep learning reconstruction technique provides a geometrically agnostic reconstruction where an analytic likelihood does not exist.  We provide an example analysis of detector performance in ton-scale detector by reconstructing a $^{90}Y$ $\beta$ emitter point source as a function of WbLS concentration.  We enforce a lower bound on the energy and only look at events with an energy higher than 1.5 MeV to remove resolution effects due to efficiency.  The results of the reconstruction are shown in~\cref{fig:tonscale_reco}.  One sees that, as the fraction of scintillator in the detector is increased, the vertex and energy resolution is improved, while the directional resolution degrades.

For this example study and others performed on ton-scale detectors, we were able to use identical network architecture and training hyperparameters.  In principle, one can also include particle identification directly in this reconstruction, by adding an additional parameter in the hypothesis variable $\params$ that signifies particle type (\ie, $e^-$, $e^+$, $\gamma$ or $n$).  Once this is accomplished, a multi-particle reconstruction can be performed.  For example, extracting information from both the $e^+$ and $n$ in an inverse beta decay can provide additional information about the incoming antineutrino.

The number of hits as well as where they appear in the detector strongly varies from interaction to interaction. The machine learning method has to be flexible about its input, especially when performing an analysis over various concentrations of scintillating material.  While varying WbLS concentration produces a trade-off in resolution, there are a number of design features that can be tuned to increase performance more generally.  Some examples include increasing the photocathode coverage, decreasing the size of the photosensor, and adding dichroicons~\cite{Kaptanoglu:2019gtg} that can provide useful wavelength-dependent separation of Cherenkov and scintillation light.  This reconstruction technique does not require special tuning to accept inputs from any of these design changes and streamlines the process for a cost/benefit analysis for various design choices during the R\&D phase of development.

\section{Discussion}

\subsection{Strengths \& Weaknesses of Approach}

To put our proposed method into context, we will discuss a few strengths and weaknesses with respect to other methods.
The canonical way of applying ML for reconstruction is in the configuration of a regressor, where direct estimates of the quantities of interest are returned. This, however, requires an architecture to input all features at once, which can lead to relatively large networks, and complex architectures to embed the data. The correct embedding can in particular be a challenge for detector geometries that do not easily map onto inputs of successful NN architectures, such as CNNs that require data in the form of images. Furthermore, a regressor requires domain-specific definitions of the loss function, and depending on this choice the results will differ.\footnote{For example: Should an energy estimator be trained in the linear or logarithmic space? Should the errors be penalized linearly (L1 loss), quadratically (L2 loss), or some other choice?}
At the same time, a regressor is comparatively fast for parameter inference.
In contrast, our proposed method offers higher flexibility, access to a likelihood, and in the \perdom formulation a geometry-agnostic architecture and training.
Having a likelihood at hand, one has in principle the full information of the relationship between the data and the model, and any standard inference machinery can be used. Traditionally, such likelihood functions had to be designed and implemented ``by hand''. With our method, this difficult task can be delegated to the computer that simply needs to be provided a forward simulation. In the \perdom configuration it can also be used for different detector geometries without re-training, enabling rapid turnaround design studies and accurate design optimization.
A possible downside of any ML approach, in contrast to traditional methods, is the need for relatively large sets of generated events to train the models.

\subsection{Conclusions and Outlook}

Under the assumption of independence, the likelihood for event reconstruction in multi-sensor detectors can be broken apart into smaller terms in different ways. Although mathematically equivalent, a salient difference between our \perdom and \alldom formulations is the number of terms that explicitly need to be evaluated and summed over to obtain the full likelihood or its functional equivalent.
In either case, we show how neural networks can be trained on forward simulation to approximate the different terms using the ratio estimator. The resulting values are no longer the likelihood itself, but a quantity proportional to it, which can be used unchanged for the task of parameter inference.
We have illustrated and validated our technique by comparing it with the analytically known true likelihood in a toy physics setup. Our method attains nearly optimal performance, and the usage of different inference techniques was demonstrated.
An interesting application of our technique is the optimization of detector properties as a function of estimated reconstruction resolutions. We illustrated this with a toy model with varying the sensor density, revealing useful relationships between sensor spacing and parameter resolutions.
Finally, we applied our method using a realistic particle physics simulation. The resolutions in energy, angle and vertex position were assessed as a function of the concentration of the water based liquid scintillator.

The flexibility of our method makes it an interesting candidate to serve as event reconstruction for future neutrino telescopes. For instance, the IceCube Upgrade \cite{Ishihara:2019aao} will feature segmented modules at different positions within non-homogeneous ice and with PMTs pointing at various angles---a problem that is naturally embedded in our formulation. Similarly, deep-sea neutrino telescopes, such as KM3NeT~\cite{KM3Net:2016zxf} also feature multi-PMT modules, and in addition their geometry is not fixed, having free floating strings in the water, for which our \perdom formulation offers a solution that can be applied to changing geometries.
Additionally, having a function proportional to the likelihood available allows for a consistent treatment of uncertainties of parameters of interest (\eg, the uncertainty of the direction pointing back to possible astrophysical neutrino sources), important to the scientific data analyses.
Furthermore, our method is already finding application in the Eos detector~\cite{Klein:2022tqr} design process, and could readily be used to reconstruction data produced by other tank-based water Cherenkov detectors (see, \eg,~\cite{Super-Kamiokande:2002weg,Theia:2019non}).
Finally, even beyond neutrino detectors, the generality of our approach should allow for it to be used in most multi-channel detectors.

\clearpage

\section*{Acknowledgments}

The authors are grateful to T. Kaptanoglu for providing RAT-PAC simulated data and guidance on how to use them.

The work of G.W. was supported (in part) by the Consortium for Monitoring, Technology, and Verification under the Department of Energy National Nuclear Security Administration award number DE-NA0003920.  
Elements of the work by D.F.C. and G.W. were performed under the auspices of the U.S. Department of Energy by Lawrence Livermore National Laboratory under Contract DE-AC52-07NA27344.

The work of J.W. and S.B. has been supported by the Cluster of Excellence “Precision Physics, Fundamental Interactions, and Structure of Matter” (PRISMA+ EXC 2118/1) funded by the German Research Foundation (DFG) within the German Excellence Strategy (Project ID 39083149)

The work of P.E. was supported by the Deutsche Forschungsgemeinschaft (DFG, German Research Foundation) under Germany's Excellence Strategy – EXC-2094 – 390783311 and the Sonderforschungsbereich (Collaborative Research Center) SFB1258 ‘Neutrinos and Dark Matter in Astro- and Particle Physics’.

\section*{Author contributions}

P.E. conceived of the original idea, P.E., A.T.F. and J.W. developed the statistical aspects and the technical implementations.  P.E., A.T.F. and J.W. ran simulations, collected data, and contributed to the developments of the neural networks. J.W. optimized many aspects of the neural networks, their training and validation. G.W. applied the technique to the ton-scale detector and performed the corresponding studies. D.F.C. is the supervisor of G.W., and S.B. is the supervisor of J.W. The project leader was P.E. All authors regularly discussed the work and contributed to the final manuscript.

\bibliography{references}

\appendix

\section{EML formulations}

\label{sec:EML_appendix}

In the case of a single sensor, the extended likelihood function takes the following form:  
\begin{equation}
    \mathcal{L}(\params|\obs) = \left[\prod_{i=1}^{N} p(\obs_i | \params)\right] \mathcal{P}(N | \Lambda(\params)),
\end{equation}
where $p(x|\params)$ is a \ac{PDF} in the quantity $x$ observed by the sensor, and $\Lambda(\params)$ is the expected number of hits given the hypothesis \params, and $\mathcal{P}$ denotes the standard Poisson distribution. The \ac{EML} estimator $\widehat{\params}$ is then the value of \params that maximizes $\mathcal{L}$.

Multiple sensors can be accommodated by combining the likelihood $\mathcal{L}_s$ for each individual sensor,
\begin{equation}
    \label{eq:dom_formulation_appendix}
    \mathcal{L}(\params|\obs) = \prod_{s=1}^{N_{\mathrm{sens}}} \mathcal{L}_s(\params|\obs) = \prod_{s=1}^{N_{\mathrm{sens}}}\left[\prod_{i=1}^{N_s} p_s(\obs_{i, s} | \params) \right] \mathcal{P}(N_s | \Lambda_s(\params)),
\end{equation}
where $s$ is a sensor index and $N_{\mathrm{sens}}$ is the total number of sensors. All quantities have acquired $s$ subscripts because the hit PDF, expected number of hits, and observed number of hits varies between sensors. The above treatment assumes that the statistical variations for each sensor are independent, which is typically the case. 

\cref{eq:dom_formulation_appendix} has the practical disadvantage that, while there may be only a small number of hits in a given event, one must determine $\Lambda_s$ for every sensor, including those without hits. This can become quite computationally expensive in cases where there are many thousands of sensors. 
To address this disadvantage, we rewrite \cref{eq:dom_formulation_appendix} as
\begin{equation}
    \begin{split}
        \mathcal{L}(\params|\obs) & = \prod_{s=1}^{N_{\mathrm{sens}}}\left[\prod_{i=1}^{N_s} p_s(\obs_{i, s} | \params) \right] \Lambda_s(\params)^{N_s} e^{-\Lambda_s(\params)} \\
        & = \prod_{s=1}^{N_{\mathrm{sens}}}\left[\prod_{i=1}^{N_s}\Lambda_s(\params) p_s(\obs_{i, s} | \params) \right] e^{-\Lambda_s(\params)} \\
        & = \left[\prod_{i=1}^{\sum_{s=1}^{N_{sens}}N_{s}}\Lambda_{s_i}(\params) p_{s_i}(\obs_{i} | \params) \right] e^{-\sum_{s=1}^{N_{sens}}\Lambda_{s}} \\
        & = \left[\prod_{i=1}^{N_{\mathrm{tot}}}\Lambda_{s_i}(\params) p_{s_i}(\obs_{i} | \params) \right] \frac{\Lambda_\mathrm{tot}(\params)^{N_{\mathrm{tot}}}}{\Lambda_\mathrm{tot}(\params)^{N_{\mathrm{tot}}}} e^{-\Lambda_\mathrm{tot}(\params)} \\
        & = \left[\prod_{i=1}^{N_{\mathrm{tot}}} \frac{\Lambda_{s_i}(\params)}{\Lambda_\mathrm{tot}(\params)} p_{s_i}(\obs_i | \params) \right] \Lambda_\mathrm{tot}(\params)^{N_{\mathrm{tot}}} e^{-\Lambda_\mathrm{tot}(\params)}
    \end{split}
\end{equation}

Here there is a term for each observed hit, with $s_{i}$ indicating the sensor for hit $i$, and an overall factor relating the total number of observed hits, $N_{\mathrm{tot}} = \sum_{s=1}^{N_{\mathrm{sens}}}N_s$, to its expectation. In this formulation, non-hit sensors contribute solely through $\Lambda_{\mathrm{tot}}$. This paper presents and characterizes a technique for calculating $\Lambda_\mathrm{tot}(\params)$ that does not involve explicit iteration over all sensors.  

Finally, introducing $q_{s}(\obs | \params) \coloneqq \frac{\Lambda_{s}(\params)}{\Lambda_\mathrm{tot}(\params)} p_{s}(\obs | \params)$, we arrive at:
\begin{equation}
   \mathcal{L}(\params|\obs)  = \left[\prod_{i=1}^{N_{\mathrm{tot}}} q_{s_i}(\obs_i | \params) \right] \mathcal{P}(N_{\mathrm{tot}} | \Lambda_\mathrm{tot}(\params)).
\end{equation}.

\section{Toy Experiment}
\label{sec:toy_appendix}
As a test bed for our method, we define a close-to-realistic toy detector model that still retains the ability to evaluate likelihoods analytically. The availability of the true likelihood is of essence for our validation and comparisons, while in most real experiments such a true likelihood is not available.

We express the detector response $x$ in terms of a set of event parameters \params with a segmented interaction model and parameterized time and distance functions.
The event model itself consists of eight parameters: vertex position $V_{xyz}$, interaction time $t$, cascade energy $E_{cscd}$, and a track with energy $E_{trck}$ and direction $(\vartheta, \varphi)$. The cascade part is modeled as an isotropic light emission with 12819 Cherenkov photons per GeV \cite{Aartsen:2013rt}. The track part is modeled as a line of cascade emissions placed every 1\,m along the track, to a total length of 4.5\,m/GeV, and a constant 2451 photons of Cherenkov light emitted each meter \cite{Aartsen:2013rt}.

\begin{figure}
    \centering
    \includegraphics[width=0.6\textwidth]{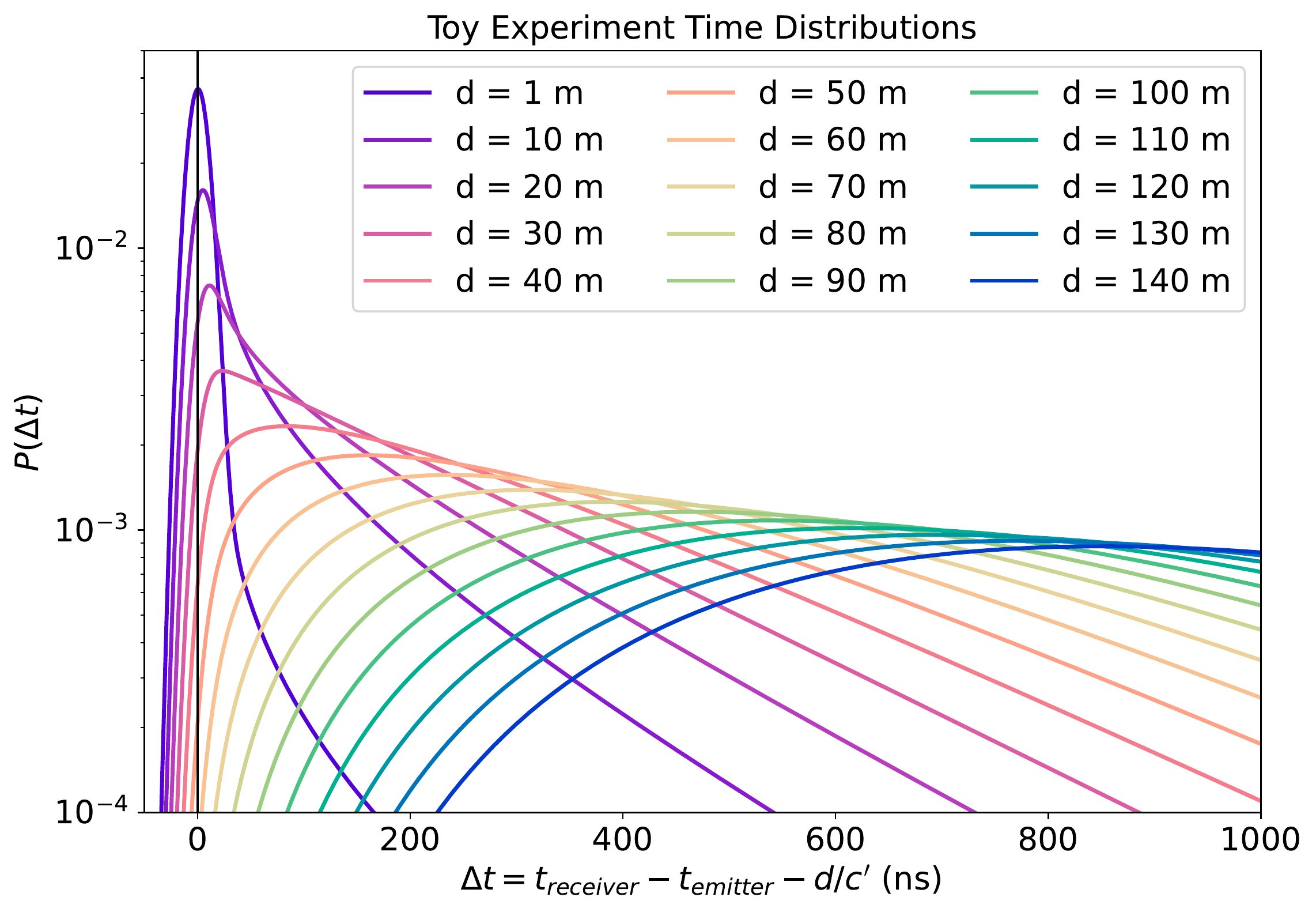}
    \caption{Gauss convolved Pandel functions for a few choices of distance $d$, and distribution parameters fixed at the values indicated in the text.}
    \label{fig:pandel}
\end{figure}

The amount of photons arriving at a detector element scales as $1/r^2$ geometrically, plus and additional absorption of $e^{-r/\lambda_a}$, where we set the absorption length $\lambda_a=100$\,m.
The time arrival distribution at sensors is modeled via a convolved Pandel function \cite{vanEijndhoven:2007ge}, for which we set the absorption length to the same value $\lambda_a=100$\,m, the scattering length $\lambda_s=30$\,m, assume a refractive index $n=1.3$, and the coefficient $\tau=500$. The Gaussian smearing of the convolution is set to 10\,ns. \Cref{fig:pandel} shows the resulting time distributions for a few different emitter-receiver distances $d$. These timing distributions model the residual time $\Delta t$ of photon detection, i.e. the time relative to the emission time minus the time of the shortest path $d/c'$.

\section{Implementation details}
\label{sec:imp_details}
For our toy models we use $\mathcal{O}(10^{6})$ events uniformly simulated over all \params parameters of interest. 90 percent of the events are used to train the networks while 10 percent remain as validation set. The training proceeds until the validation loss stagnates, which usually takes around 7-15 epochs for a \hitnet and 100 epochs for a \chargenet in the \alldom formulation while less than 10 epochs for a \chargenet in the \perdom formulation, respectively. On four GPUs (Nvidia GTX 1080ti \cite{GPU}) this takes $\mathcal{O}$(hrs). The \emph{adam} optimizer \cite{adam} is used for training with a learning rate set to $10^{-4}$, and a batch size of 4096 was used.\\
The dense NN part of the model consists of 14 hidden layers with 250 neurons each (\hitnet in both formulations and \chargenet in \perdom formulation), or 13 hidden layers with 150 neurons each (\chargenet in \alldom formulation), respectively. The \emph{swish} \cite{swish} activation function is used in all hidden layers.\\

\begin{table}[h]
\centering
\begin{tabular}{p{1.5cm}||p{5cm}|p{5cm}}
  & \hitnet & \chargenet \\
\hline
\hline
 \obs & position ($x, y, z$), and time ($t$) of hit & number of hits in sensor (\perdom) or detector (\alldom) \\
\hline
 \params & position ($x, y, z$), time ($t$), angle ($\varphi, \vartheta$), energy ($E$), and inelasticity ($I$) of source &

 position ($x, y, z$), angle ($\varphi, \vartheta$), energy ($E$), and inelasticity ($I$) of source \\
\end{tabular}
\end{table}

\end{document}